\documentclass[10pt,conference]{IEEEtran}

\usepackage[utf8]{inputenc}
\usepackage[T1]{fontenc}
\usepackage{lmodern}

\usepackage[DIV=14,BCOR=2mm,headinclude=true,footinclude=false]{typearea}

\usepackage{url}
\usepackage{amsmath}
\usepackage{amssymb}
\usepackage{amsthm}
\usepackage{dsfont}
\usepackage{mathtools}
\usepackage{amsfonts}
\usepackage{siunitx}

\usepackage[lined,ruled,noend,linesnumbered,onelanguage,norelsize]{algorithm2e}

\usepackage{graphicx}
\usepackage{subfigure}

\usepackage{tikz}
\usepackage{tkz-graph}

\usepackage{ytableau}

\usepackage{gensymb}

\usepackage[format=hang,justification=centering,font=small,labelfont=bf,singlelinecheck=off]{caption}

\usepackage{supertabular}
\usepackage{multirow}
\usepackage{booktabs}
\usepackage{tabularx}
\usepackage{enumitem}

\usepackage{todonotes}

\usepackage[breaklinks,colorlinks,citecolor=blue,linkcolor=blue,
  pdftitle={Ambiguities in DoA Estimation with Linear Arrays},
  pdfauthor={Frederic Matter, Tobias Fischer, Marius Pesavento, Marc Pfetsch}]{hyperref}

\theoremstyle{plain}
\newtheorem{lemma}{Lemma}[section]
\newtheorem{theorem}[lemma]{Theorem}

\newtheorem{proposition}[lemma]{Proposition}

\theoremstyle{definition}
\newtheorem{definition}[lemma]{Definition}
\newtheorem{example}[lemma]{Example}
\newtheorem{remark}[lemma]{Remark}

\newcommand{\g}{\raisebox{0.25ex}{\tiny $>$}}

\newcommand{\define}{\coloneqq}
\newcommand{\enifed}{\eqqcolon}

\newcommand{\rank}{\operatornamewithlimits{rank}}

\newcommand{\suchthat}{\,:\,}
\newcommand{\R}{\mathds{R}}
\newcommand{\N}{\mathds{N}}
\newcommand{\Z}{\mathds{Z}}
\newcommand{\C}{\mathds{C}}

\newcommand{\Q}{\mathds{Q}}

\newcommand{\T}{^\top}

\newcommand{\card}[1]{\lvert{#1}\rvert}
\newcommand{\abs}[1]{\lvert{#1}\rvert}

\newcommand{\modulo}{\operatornamewithlimits{mod}}

\newcommand{\bA}{{\bf A}}

\newcommand{\bP}{{\bf P}}
\newcommand{\bw}{{\bf w}}
\newcommand{\bb}{{\bf b}}
\newcommand{\by}{{\bf y}}
\newcommand{\bz}{{\bf z}}

\newcommand{\bx}{{\bf x}}
\newcommand{\br}{{\bf r}}
\newcommand{\ba}{{\bf a}}
\newcommand{\bq}{{\bf q}}
\newcommand{\bof}{{\bf f}}
\newcommand{\bV}{{\bf V}}
\newcommand{\bv}{{\bf v}}
\newcommand{\bB}{{\bf B}}

\newcommand{\bX}{{\bf X}}

\newcommand{\bTheta}{{\boldsymbol \Theta}}
\newcommand{\blambda}{{\boldsymbol \lambda}}
\newcommand{\bdelta}{{\boldsymbol \delta}}
\newcommand{\balpha}{{\boldsymbol \alpha}}
\newcommand{\bbeta}{{\boldsymbol \beta}}
\newcommand{\bsigma}{{\boldsymbol \sigma}}

\newcommand{\bPhi}{{\boldsymbol \Phi}}

\newcommand{\Real}{\operatornamewithlimits{Re}}
\newcommand{\Imag}{\operatornamewithlimits{Im}}
\newcommand{\e}{{\operatornamewithlimits{e}}}

\IEEEoverridecommandlockouts

\begin{document}
	
	\title{Ambiguities in Direction-of-Arrival Estimation\\ with Linear Arrays}
	\author{Frederic Matter, Tobias Fischer, Marius Pesavento, Marc E. Pfetsch
		\thanks{This work was supported by the EXPRESS II project within the DFG
			priority program CoSIP (DFG-SPP 1798).}}
	\date{\today}
	\maketitle

\begin{abstract}
In this paper, we present a novel approach to compute ambiguities in thinned uniform linear arrays, i.e., sparse non-uniform linear arrays, via a mixed-integer program. Ambiguities arise when there exists a set of distinct directions-of-arrival, for which the corresponding steering matrix is rank-deficient and are associated with nonunique parameter estimation. Our approach uses Young tableaux for which a submatrix of the steering matrix has a vanishing determinant, which can be expressed through vanishing sums of unit roots. Each of these vanishing sums then corresponds to an ambiguous set of directions-of-arrival.
We derive a method to enumerate such ambiguous sets using a mixed-integer program and present results on several examples.
\end{abstract}

\section{Introduction}
\label{sec:introduction}

\noindent
Sparse uniform linear arrays, obtained from thinning uniform linear arrays (ULAs), are widely used in sensor array processing due to their ability to maintain the aperture size of the corresponding full ULA while reducing the number of array elements. This is associated with as significant decrease of the array costs including power consumption, hardware and computational complexity \cite{Oliveri_2009} as well as a reduction in the mainbeam width and mutual coupling \cite{Liu_2016a,Liu_2016b}. A variety of array thinning techniques have been proposed in literature to control the side and grating lobe levels of the array. These techniques can be classified into three main areas \cite{Toso_2007}:  i) \emph{analytic thinning}, e.g., based in prime number selection, where the array is formed from a $\lambda/2$ ULA by selecting sensors at prime multiples of the baseline as, e.g., in nested arrays, coprime arrays, and minimum redundancy arrays \cite{Pal_2010,PP_2011,Moffet_68}, ii) \emph{statistical thinning} techniques, where sensors are selected randomly \cite{Rocca_2011}, and iii) \emph{optimization based thinning} in which an appropriate error function is minimized \cite{Wang_2018,Boyd_2009,Haupt_94,Isemia_2004,Robinson_2004,Murino_96}. 

Low side and grating lobe levels are important indicators for the resolution performance for sparse linear arrays in beamforming and Direction-of-Arrival (DoA) estimation applications. However, a fundamental question in the context of multi-source DoA estimation in thinned arrays is that of the maximum number of sources that can uniquely be determined from an array measurement and the characterization of ambiguities in the measurements. Ambiguities in the array manifold arise when there exists a set of distinct directions-of-arrivals in the field of view, for which the corresponding steering matrix is rank-deficient. In this case it is impossible to uniquely determine the parameters of interest from single snapshot measurements.

The concept and mathematical framework for the ambiguity problem in
direction finding for linear arrays dates back to Schmidt~\cite{Schmidt81},
who introduced the rank of an ambiguity based on linear dependency in the
array steering matrix and thus classified ambiguities based on this notion
of rank. In the subsequent years, research focused onto specific array
geometries which have desirable properties or are free of certain types of
ambiguities~\cite{LoMarple92,TanGohTan96,TanGoh94,TanOE93}.
For the popular ULA geometry it is known that no
ambiguities exist, if the intersensor spacing is at most half wavelength~\cite{ManikasP98,
	Skolnik80, TanBeachNix02, TanGohTan96}. A linear array that is not
uniform is called a \emph{non-uniform} linear array.

Manikas and Proukakis~\cite{ManikasP98} use tools from differential
geometry in order to analyze ambiguities for general linear arrays.
They derive a sufficient condition for the presence of ambiguities in a
linear array. This sufficient condition implies that
every non-uniform linear
array with integer positions suffers from ambiguities,
see~\cite{AbramovichSpencerGorokhov99}. Moreover, Manikas and Proukakis
derive a class of ambiguities that are present in every non-uniform integer
linear array, see~\cite[Theorem~2.2]{ManikasP98}. We call these ambiguities
\emph{uniform ambiguities}, since they are derived from a uniform
partitioning of the array manifold.

For symmetric linear arrays, which can be shifted
globally to positions $r_1, \dots, r_M$, such that $\sum_{i=1}^M (r_i)^n = 0$ holds for all odd
$n \in \N$, Manikas and Proukakis identify additional ambiguities,
see~\cite[Theorem~3]{ManikasP98}. This ambiguity criterion for symmetric
linear arrays is generalized in Dowlut~\cite[Theorem~3]{Dowlut02} and
Manikas~\cite[Theorem~7.1]{Manikas04} to a whole class of so-called
\emph{non-uniform} ambiguities that depend on a parameter
$n \in \N$.
Moreover,
Manikas~\cite[Lemma~7.1]{Manikas04} states that these non-uniform
ambiguities converge to a uniform ambiguity for $n \to \infty$.

Wax and Ziskind~\cite{WaxZ89} derive conditions for which the signal
model~\eqref{steering_model} has a unique solution based on the number of
sensors in the array and the spark of the steering matrix.

In a recent paper, Achanta et al.~\cite{AchantaBDDJM17} investigate the
spark of DFT matrices and use vanishing sums in order to find rows of a DFT
matrix that induce a matrix with full spark. However, they do not draw the
connection to generalized Vandermonde matrices and the Schur polynomial.
Translated to our setting, the results of~\cite{AchantaBDDJM17} can determine
which linear arrays do not have any ambiguities within a fixed
set of certain DoAs. In contrast, our paper determines
ambiguities a fixed linear array.

\paragraph*{Contribution}
Which non-uniform integer linear array suffers from
ambiguities, besides the ones already known in the literature? In order to
answer this question, we need to find steering vectors $\ba(\theta_i)$
(or more precisely DoAs~$\theta_i$) such that the steering matrix for the
position vector~$\br$
has spark at most~$M$. This means that the $M$ sensors cannot uniquely
localize the signal sources with directions $\theta_i$. In the following,
we present a novel approach for identifying ambiguities in (non-uniform)
integer linear arrays. This approach can in theory be used to compute
\emph{all} ambiguities from which a given integer linear array suffers, see
Section~\ref{sec:AlgebraicRootsSchurPoly}. However, those computations are
too expensive to be of practical use, even for small arrays. By making a
(small) structural assumption, we are able to formulate a mixed-integer problem (MIP) that is
capable of enumerating ambiguities, see Section~\ref{sec:EnumMVS}. At least
for non-symmetric integer linear arrays, we show that this approach can
find all ambiguities already known in the literature (namely those
from~\cite[Theorem~2.2]{ManikasP98}), see
Proposition~\ref{prop:FindAllUniformAGS}. Moreover, we demonstrate the usefulness
of our approach in Section~\ref{sec:CompRes} by presenting examples of
non-symmetric integer linear arrays for which we find many previously
unknown ambiguities.

\begin{remark}
	Our notion of ambiguity and the failure to uniquely resolve source
	signals can also be used to characterize ambiguities in co-arrays corresponding to non-fully augmentable arrays~\cite{AbramovichSpencerGorokhov99}.
\end{remark}

In this paper, the following notations are used:
matrices are denoted by boldface uppercase letters~$\bA$, vectors
are denoted by boldface lowercase letters $\ba$, and scalars are
denoted by regular letters~$a$. 
Symbols $(\cdot)\T$ and $(\cdot)^{-1}$ denote the
transpose, and inverse of the (matrix)
argument. For~$n \in \N$, we define~$[n] \define \{1,\dots,n\}$.

\section{Problem Description}
\label{sec:problemdescription}

\noindent
Consider a sparse (thinned) uniform linear array composed of $M$ sensors located on the $x$-axis in $\R^2$ at positions corresponding to integer multiples $r_1 < r_2 < \cdots < r_M$ of a common baseline $d \in [0,1]$ measured in half wavelength. Define $\br = (r_1,\dots,r_M) \in \Z^M$.
A superposition of~$L$ narrowband waveforms emitted from 
sources at azimuth angles
$\bTheta = (\theta_1,\dots, \theta_L)$ is impinging on the array. Throughout the paper, we denote
the azimuth angles $\theta \in \bTheta$ as
\emph{Directions-of-Arrival (DoA)}. The received signal~$\by$ in a
noise-free setting is expressed as
\begin{align}
  \by = \bA(\bTheta)\, \bx,
  \label{steering_model}
\end{align}
where $\bx\in \C^L$ is the emitted signal and $\bA(\bTheta) =
[\ba(\theta_1), \dots, \ba(\theta_L)] \in \C^{M\times L}$ the
array steering matrix with columns
\begin{align}\label{eq:SteeringVector}
  \ba(\theta) = [z^{r_1}, \dots, z^{r_M}]\T,
\end{align}
for $\theta\in \bTheta$ and $z = \e^{-j \pi d \cos(\theta)}$. By using the
\emph{electrical angle}~$\Phi = -\pi d\cos(\theta)$, this simplifies
to~$z = \e^{j \Phi}$. We then denote the array steering matrix with~$\bA(\bPhi)$.

In the case of linear measurement systems, unique recovery is assessed from the \emph{spark} of the measurement matrix~$\bA$, which is defined as the smallest number of linearly dependent columns in~$\bA$. If the spark is large enough, then it is possible to
uniquely recover~$\bx$ from its measurements~$\bA\bx$, even if the linear
system is underdetermined, see, e.g., the book~\cite{FoucartRauhut13}.

Uniqueness of the measurement model~\eqref{steering_model} is
of importance for DoA estimation as well and directly related to the number of sources that can be identified from the measurements~\cite{WaxZ89}. In our setting, the steering
matrix~$\bA(\bTheta)$ depends on the unknown directions from which the
signals in~$\bx$ impinge on the linear array~$\br$. This leads to a
generalized notion of the spark: The ability to uniquely recover a
signal~$\bx$ coming from directions~$\bTheta$ depends on the spark of the
induced steering matrix~$\bA(\bTheta)$. If the spark of~$\bA(\bTheta)$ for a
fixed linear array~$\br$ is not full, then there exists a subset of the
columns of~$\bA(\bTheta)$, i.e., a subset of steering vectors, which are
linearly dependent and thus induce a rank-deficient submatrix
of~$\bA(\bTheta)$. This is called an \emph{ambiguity}, which can be formally
defined as follows.

\begin{definition}[\cite{ManikasP98}]\label{def:ambiguoussetarclengths} Let
  $\br \in \R^M$ be a linear array with $M$ sensors and let
  $\bTheta = [\theta_1,\dots,\theta_L]\T$ be an ordered vector of DoAs with
  $L \leq M$. Then $\bTheta$ is called an \emph{ambiguous vector of DoAs}, if
  \begin{align*}
    \rank(\bA(\bTheta)) = \rank\big([\ba(\theta_1),\dots,\ba(\theta_L)]\big) < L.
  \end{align*}
  Furthermore, its \emph{rank of ambiguity} is defined as
  $\rho_a = \rank(\bA(\bTheta)) \in \N$.
\end{definition}

An important question is whether the steering matrix~$\bA(\bTheta)$
for a fixed linear array and a vector of directions~$\bTheta$ has full
spark, i.e., if the spark is given by~$\min\,\{M,L\}$.


Note that more signal sources than sensor positions~$M$ always produce an ambiguity, since the rank of
the corresponding submatrix can be at most $M$. Thus, we focus on the
search of ambiguities with at most $M$ signal sources.

In order to simplify the presentation, we make some assumptions without
loss of generalization.

\begin{enumerate}[label=(A\arabic*),series=assump]
\item\label{assump:FixFirstSensorPosition} We assume that the first sensor of a linear array
  $\br = (r_1,\dots,r_M)\T$ is located in the origin, i.e., $r_1 = 0$,
  since a global shift of the array positions does not change the
  ambiguities.
\item\label{assump:BoundsTheta} We assume that the electrical angles~$\Phi$
  lie in $\Omega = [-\pi d,\pi d]$, since $\ba(\theta) =
  \ba(2\pi - \theta)$ for $\theta \in [0,2\pi]$. For~$d=1$, we assume~$\Phi \in [-\pi, \pi)$,
  since in this case~$\ba(0) = \ba(\pi)$. This implies a one-to-one correspondence
  between electrical angles and DoAs, so that from now on, we will also use
  electrical angles.
\item \label{assump:FixFirstDoA} An appropriate global rotation of the DoAs
  amounts to multiplying the steering matrix~$\bA(\bTheta)$ with a constant
  diagonal matrix with unit complex entries, which retains the ambiguity
  property. Thus, we assume without loss of generality that~$\theta_1 = 0$ resulting in an electrical angle
  $\Phi_1 = -\pi d$.
\end{enumerate}

Furthermore, we assume that $\br \in \N^M$. Recall that~$\br$ denotes the
integer multiples of the common baseline~$d \in [0,1]$, measured in half
wavelength.  Such sparse uniform sampling is often used in practical
systems due the simplicity of the hardware design and several attractive
features associated with it, such as the existence of efficient search-free
rooting-based DoA estimation
techniques~\cite{Haardt14,Barabell83,Rao88,Li98,Stoica90} and virtual
signal decorrelation procedures~\cite{Shan85,Thakre09,Pesavento00}.
  

\section{A novel approach for detecting Ambiguities in Linear Arrays}
\label{sec:ambiguityapproach}

\noindent
Let
Assumptions~\ref{assump:FixFirstSensorPosition}--\ref{assump:FixFirstDoA}
hold and let~$\br \in \N^M$. We first give a compact overview
over our procedure to find ambiguities that is presented in the next
sections.

\begin{enumerate}[leftmargin=3ex]
\item The array steering matrix $\bA(\bPhi)$ for $\br$ induces ambiguities if
  some of its $M\times M$ submatrices have a nonzero
  determinant. Therefore, we search for $M$ electrical angles
  $\Phi_1, \dots, \Phi_M$ such that the determinant of their induced
  submatrix of $\bA(\bPhi)$ vanishes.
\item Every submatrix of $\bA(\bPhi)$ is a generalized Vandermonde determinant,
  see Section~\ref{sec:vandermonde}. This determinant is divisible by the
  classical Vandermonde determinant, and this quotient is equal to the
  Schur polynomial $s_\blambda(\bz)$, see
  Equation~\eqref{eq:SchurPoly}. Thus, instead of searching for roots of
  the generalized Vandermonde determinant, we can search for roots of the
  Schur polynomial.
\item The Schur polynomial $s_\blambda(\bz)$ can be represented using
  semistandard Young tableaux (SSYTs), see Section~\ref{sec:youngtableaux}
  and Lemma~\ref{Lem_YoungRule} therein.
\item In our case, $z_i = \e^{j \Phi_i}$ for
  $i \in [M]$, and if $\Phi_i \in \{2\pi\,q
  \suchthat q\in \Q\}$, the Schur polynomial is a sum of unit
  roots. This means, we search unit roots such that their sum vanishes and
  the relationship to the SSYTs due to Lemma~\ref{Lem_YoungRule} is satisfied, see
  Section~\ref{sec:vansums}.
\item We construct these vanishing sums of unit roots by adding up rotated minimal
  vanishing sums. This can be formulated as a mixed-integer
  problem, see Section~\ref{sec:EnumMVS}.
\end{enumerate}

\subsection{Generalized Vandermonde Matrix}
\label{sec:vandermonde}

\noindent
Our goal is to determine whether there exist electrical angles $\Phi_1 < \cdots < \Phi_M$ and
an $M \times M$ submatrix of the corresponding array steering matrix for an integer
linear array $\br$ with zero determinant. Every $M\times M$ submatrix of $\bA(\bPhi)$ is
a \emph{generalized Vandermonde matrix} of the form
\begin{align*}
  \bB_\br(\bz) =
  \begin{pmatrix}
    z_1^{r_1} & \dots & z_M^{r_1} \\
    z_1^{r_2} & \dots & z_M^{r_2} \\
    \vdots & \ddots & \vdots \\
    z_1^{r_M} & \dots & z_M^{r_M}
 \end{pmatrix},
\end{align*}
with $z_i = \e^{j \Phi_i}$, $i\in [M]$ and
$\Phi_{1} < \cdots < \Phi_{M}$. We define the
polynomial
\begin{align}\label{eq:GVD}
  V_\br: \C^M\rightarrow \C, \quad  V_\br(\bz)\define \det(\bB_\br(\bz))
\end{align}
as the generalized Vandermonde determinant.
Every root of $V_\br$ induces an ambiguity.

If $\br = (0, 1, \dots, M-1)\T$, then $V_\br: \C^M\rightarrow \C$ is the
classical Vandermonde determinant
\begin{align*}
  V(\bz) = \prod\limits_{1\leq i < k \leq M} (z_k - z_i).
\end{align*}
In this case, we obtain that $V(\bz) = 0$ if and only if $z_i = z_k$
for indices $i \neq k$. This means that ambiguities only arise if there are two equal electrical angles $\Phi_i = \Phi_k$ for
$i \neq k$~\cite{Sidiropoulos01}.

For steering matrices with a non-uniform linear array,
$\bB_{\br}(\bz)$ is a so-called \emph{generalized
  Vandermonde matrix}, and its determinant~$V_{\br}(\bz)$ is a
\emph{generalized Vandermonde determinant}. For more information on
generalized Vandermonde matrices and determinants, see,
e.g.,~\cite{Heineman29,BuckEtAl92,Ernst00,SchlickeweiViola00} as well
as~\cite{Mitchell81,Moore1896} for results in case that~$\bz \in \R^M$ and
that~$\bV_\br$ is defined over a finite field, respectively.


In the present paper, we have~$\bz\in \C^M$. In this case, it is well known that $V_\br(\bz)$ is divisible by the
(classical) Vandermonde determinant $V(\bz)$, see, e.g.,~\cite{Marchi99}. The ratio
\begin{align}\label{eq:SchurPoly}
  s_{\blambda}(\bz) \define \frac{V_{\br}(\bz)}{V(\bz)}
\end{align}
with $\blambda = \br - \bdelta$, and $\bdelta = (0, 1, \dots, M-1)\T$ is
the so-called \emph{Schur polynomial}. Here, $\blambda$ is sorted
non-decreasingly.

An ambiguous vector of electrical angles $[\Phi_1,\dots,\Phi_M]\T$ induces a
generalized Vandermonde matrix with a zero determinant, and thus forms a
root of the Schur polynomial $s_\blambda(\bz)$ with
$z_i = \e^{j \Phi_i}$, $i\in [M]$. In order to
find such roots, we introduce Young tableaux in the next section, as they
can be used to represent the Schur polynomial.

\subsection{Young Tableaux}
\label{sec:youngtableaux}

\noindent
The notations, definitions and statements in this section are taken from~\cite{Stanley97}. 

Given $\blambda = (\lambda_1,\dots, \lambda_M)\T$,
$0\leq \lambda_1\leq \dots \leq \lambda_M$, the \emph{Young diagram} of shape
$\blambda$ is defined as a collection of boxes that are arranged in~$M$
left-justified rows such that the number of boxes in row~$M-i+1$
is~$\lambda_i$. A \emph{semistandard Young tableau} (SSYT) of shape~$\blambda$ is
obtained by filling the Young diagram with positive integers $i \in [M]$
such that entries increase weakly along each row and increase strictly down
each column\footnote{Note that in general, a Young diagram can be filled
  with arbitrary positive integers $i \in \N$ to obtain an SSYT, but as it
  will become clear later in this section, in our setting, only integers
  $i \in [M]$ are allowed.}.

For example, if $\blambda = (1,3,4)\T$, we need to fill the Young
diagram with integers from the set $\{1,2,3\}$ and one possible SSYT is the
following:

\begin{figure}[h]
  \centering
  \begin{ytableau}
    1 & 1 & 2 & 3 \\
    2 & 3 & 3\\
    3
  \end{ytableau}
\end{figure}
For a given $\blambda$, we define $\mathcal{T}_{\blambda}$ as the set of all
SSYTs with entries in~$[M]$.  Each tableau
$T \in \mathcal{T}_{\blambda}$ defines a \emph{weight vector}
$\balpha(T) = (\alpha_1(T), \dots, \alpha_M(T))\T$, where~$\alpha_i(T)$ is the number of times~$i$ appears. By
definition,
$\sum_{i=1}^M\alpha_i(T) = \sum_{i=1}^M\lambda_i$ holds.
In the above example,
$(\alpha_1(T), \alpha_2(T), \alpha_3(T))\T = (2, 2, 4)\T$.

SSYTs have the beautiful property that they can be used to represent the
Schur polynomial. This is known in the literature under the term ``Young's
Rule'', and it plays a key role in our procedure, since it yields a
polynomial representation of the quotient of the generalized Vandermonde
determinant and the classical Vandermonde determinant.

\begin{lemma}[Stanley~\cite{Stanley97}
  ]
  \label{Lem_YoungRule}
  Consider $\blambda = (\lambda_1,\dots, \lambda_M)\T$ with
  $0\leq \lambda_1\leq \dots \leq \lambda_M$. Then,
 \begin{align*}
   s_{\blambda}(\bz) = \sum_{T\in \mathcal{T}_{\blambda}} \prod\limits_{i=1}^M z_i^{\alpha_i(T)}.
 \end{align*}
\end{lemma}

Lemma~\ref{Lem_YoungRule} is the reason why we only allowed integers
$i \in [M]$ to appear in an SSYT in the definition at the beginning of this
section. From Lemma~\ref{Lem_YoungRule}, we deduce that the degree
of~$s_{\blambda}$ is $\sum_{i=1}^M \lambda_i$. Moreover, if
$\blambda \neq 0$, then~$s_{\blambda}$ is homogeneous. For a formula of
the total number of SSYTs of shape $\blambda$ with entries in
$[M]$, see~\cite[Theorem~6.3]{FultonHarris13}, which is a
simple consequence of~\cite[Lemma~7.21.1 and Theorem~7.21.2]{Stanley97}.
For all linear arrays with positions in $(0,1,2,3,4,5)\T$,
Table~\ref{tab:SSYTs} lists the number~$N$ of corresponding SSYTs.

\begin{table}[t]
  \centering
  \caption{Number~$N$ of SSYTs for selected linear arrays.}
  \label{tab:SSYTs}
  \begin{tabular}{@{}l@{\;}r|l@{\;}r|l@{\;}r|l@{\;}r@{}}
    \toprule
          Array & $N$ &        Array  & $N$ &        Array  & $N$ &          Array  & $N$ \\
    \midrule
    $(0,2)$   &   2 &   $(0,2,3)$ &   3 & $(0,1,2,5)$ &  10 &   $(0,3,4,5)$ &  10 \\
    $(0,3)$   &   3 &   $(0,2,4)$ &   8 & $(0,1,3,4)$ &   6 & $(0,1,2,3,5)$ &   5 \\
    $(0,4)$   &   4 &   $(0,2,5)$ &  15 & $(0,1,3,5)$ &  20 & $(0,1,2,4,5)$ &  10 \\
    $(0,5)$   &   5 &   $(0,3,4)$ &   6 & $(0,1,4,5)$ &  20 & $(0,1,3,4,5)$ &  10 \\
    $(0,1,3)$ &   3 &   $(0,3,5)$ &  15 & $(0,2,3,4)$ &   4 & $(0,2,3,4,5)$ &   5 \\
    $(0,1,4)$ &   6 &   $(0,4,5)$ &  10 & $(0,2,3,5)$ &  15 &                 &     \\
    $(0,1,5)$ &  10 & $(0,1,2,4)$ &   4 & $(0,2,4,5)$ &  20 &                 &     \\
    \bottomrule
  \end{tabular}
\end{table}

\begin{figure}[t]
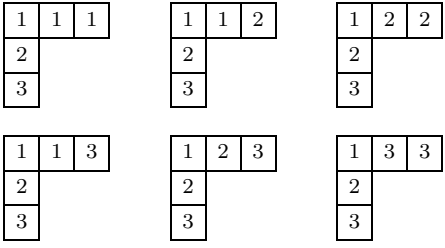

  {\footnotesize \centering
    \begin{ytableau}
      1 & 1 & 1 \\
      2 \\
      3
    \end{ytableau}
    \qquad\ 
    \begin{ytableau}
      1 & 1 & 2 \\
      2 \\
      3
    \end{ytableau}
    \qquad\ 
    \begin{ytableau}
      1 & 2 & 2 \\
      2 \\
      3
    \end{ytableau}
    \\[3ex]
    \begin{ytableau}
      1 & 1 & 3 \\
      2 \\
      3
    \end{ytableau}
    \qquad\ 
    \begin{ytableau}
      1 & 2 & 3 \\
      2 \\
      3
    \end{ytableau}
    \qquad\ 
    \begin{ytableau}
      1 & 3 & 3 \\
      2 \\
      3
    \end{ytableau}\\}
  \caption{Young tableaux for $\blambda = (1,1,3)\T$}
  \label{fig_young}
\end{figure}

\begin{example}\label{Ex_young}
  Consider the simple
  array geometry $\br = (1, 2, 5)\T$ with baseline~$d=1$. With $\br = \blambda + \bdelta$ and
  $\bdelta = (0, 1, 2)\T$, we obtain that $\blambda = (1,1,3)\T$.  Then the
  Young tableaux in $\mathcal{T}_{\blambda}$ are the ones listed in
  Figure~\ref{fig_young}. The associated Schur polynomial is
  \begin{align*}
    s_{\blambda}(\bz) &= z_1\, z_2\, z_3 \, (z_1^2 + z_1 z_2 + z_2^2 + z_1
                         z_3 + z_2 z_3 + z_3^2), 
  \end{align*}
  whose roots are all $\bz\in \C^3$ with $z_1 = \frac{1}{2}
  (-z_2-z_3\pm\sqrt{-3 z_2^2-2 z_3 z_2-3 z_3^2})$ or $z_1 = 0$ or $z_2 = 0$ or $z_3 = 0$.
\end{example}

\subsection{Algebraic root finding}
\label{sec:AlgebraicRootsSchurPoly}

\noindent
Since the Schur polynomial~$s_\blambda(\bz)$ is a complex polynomial in the
variables $z_i$, we can search roots of the Schur polynomial algebraically
by splitting each variable $z_i = x_i + j\cdot y_i$ with $x_i = \Real(z_i)$
and $y_i = \Imag(z_i)$. Thus,
$s^{\rm R}_\blambda(\bx,\by) \define \Real(s_\blambda(\bz))$ and
$s^{\rm I}_\blambda(\bx,by) \define \Imag(s_\blambda(\bz))$ are real polynomials
in the variables $x_i$, $y_i$. This yields a polynomial equation system,
which can be solved using, for instance, elimination theory or multivariate resultants. For more
details, see, e.g.,~\cite{Sturmfels2002}.

Thus, for very small examples, the roots of the Schur polynomial can be
derived algebraically.
However, if~$M$
or~$\card{\mathcal{T}_{\blambda}}$ is large, then a numerical solution is usually  difficult. Moreover, the algebraic approach becomes computationally extremely expensive.
As an alternative, the idea in Section~\ref{sec:EnumMVS} is to formulate a
mixed-integer program (MIP) whose feasible solutions correspond to roots of
the Schur polynomial. Before we state this result, we introduce
the definition of vanishing sums of unit roots in the next section, as it
turns out that these play an important role for the detection of roots of
the Schur polynomial.

\subsection{Vanishing Sums of Unit Roots}
\label{sec:vansums}
\noindent
Let $\mathcal{T}_{\blambda} = \{T_1,\dots,T_N\}$ be the set of all SSYTs of shape~$\blambda$ with
cardinality $N \define \card{\mathcal{T}_{\blambda}}$. In our
setting, the variables $z_i$ are given by $z_i = \e^{j \Phi_i}$,
$\Phi_i \in [-\pi d, \pi d]$, for $i \in [M]$. The
Schur polynomial can be written as
\begin{equation}\label{eq:schurpolyrepresentation}
  \begin{aligned}
  s_{\blambda}(\bz) 
  = \sum_{\ell=1}^N \prod\limits_{i=1}^M z_i^{\alpha_{i\ell}}
  = \sum_{\ell=1}^N \prod\limits_{i=1}^M \e^{j \alpha_{i\ell} \Phi_i}\\
  = \sum_{\ell=1}^N \e^{j \sum_{i=1}^M\alpha_{i\ell} \Phi_i},
  \end{aligned}
\end{equation}
where $\balpha = (\alpha_{1,\ell},\dots,\alpha_{M,\ell})\T \in \Z^{M\times N}$ is the vector $\balpha(T_\ell)$ defined
in Section~\ref{sec:youngtableaux}.  Recall that our goal is to search for
roots of the Schur polynomial, which means to check whether there exists a
vector $\bPhi\in [-\pi d, \pi d]^M$ such that
\begin{align}
  s_{\blambda}(\bz) &= \sum_{\ell=1}^N \e^{j \sigma_\ell} =
                       0, \label{eq:ConnectionUnitrootsYT1} \\
  \sigma_\ell &= \sum_{i=1}^M\alpha_{i\ell}\, \Phi_i\ \ (\modulo 2\pi),
                \quad \forall \ell \in [N]. \label{eq:ConnectionUnitrootsYT2}
\end{align}
Since $\e^{j\sigma_\ell} = \e^{j(\sigma_\ell + k \cdot 2\pi)}$ for
$k \in \N$, Equation~\eqref{eq:ConnectionUnitrootsYT2} only needs to hold
modulo $2\pi$.

For~$m \in \N$, define~$\omega_m = \e^{j 2\pi/m}$. Then, $\omega_m^v$ is
called an $m$-th \emph{unit root} for $v \in \{0,1,\dots,m-1\}$.
If all $\sigma_\ell / 2\pi$ are rational, then the sum
$\sum_{\ell=1}^N \e^{j \sigma_\ell} = 0$ is called a \emph{vanishing sum of
  unit roots}. 

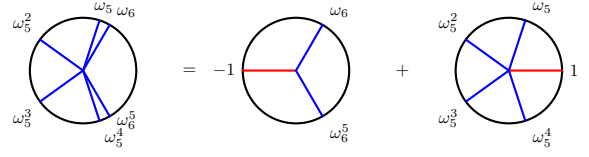
\begin{figure}[tb]
  \centering
  \begin{tikzpicture}[cap=round,>=latex,every node/.style={scale=0.65},scale=0.7]
    
    \draw[thick] (-4cm,0cm) circle(1cm);
    \coordinate (s0) at (-4,0);
    \node[above left] at (-4,0) {};
    \draw[thick, blue] (s0)--++(60:1) node[black,above right = 0.1ex and 0.1ex] {$\omega_6$};
    \draw[thick, blue] (s0)--++(-60:1) node[black,below right = -1ex and 0.1ex] {$\omega_6^5$};
    \draw[thick, blue] (s0)--++(72:1) node[black,above right = 0.1ex and -1 ex] {$\omega_5$};
    \draw[thick, blue] (s0)--++(144:1) node[black,above left = 0.1ex and 0.1ex] {$\omega_5^2$};
    \draw[thick, blue] (s0)--++(216:1) node[black,below left = 0.1ex and 0.1ex] {$\omega_5^3$};
    \draw[thick, blue] (s0)--++(288:1) node[black,below right = 0.1ex and -0.1ex] {$\omega_5^4$};
    \node[] at (-2,0) {$=$};
    \draw[thick] (0cm,0cm) circle(1cm);
    \coordinate (s) at (0,0);
    \node[above left] at (0,0) {};
    \draw[thick, red] (s)--++(180:1) node[black,left = 0.1ex] {$-1$};
    \draw[thick, blue] (s)--++(60:1) node[black,above right = 0.1ex and 0.1ex] {$\omega_6$};
    \draw[thick, blue] (s)--++(-60:1) node[black,below right = 0.1ex and 0.1ex] {$\omega_6^5$};    
    \node[] at (2,0) {$+$};
    \draw[thick] (4cm,0cm) circle(1cm);
    \coordinate (s2) at (4,0);
    \node[above left] at (4,0) {};
    \draw[thick, red] (s2)--++(0:1) node[black,right = 0.1ex] {$1$};
    \draw[thick, blue] (s2)--++(72:1) node[black,above right = 0.1ex and 0.1ex] {$\omega_5$};
    \draw[thick, blue] (s2)--++(144:1) node[black,above left = 0.1ex and 0.1ex] {$\omega_5^2$};
    \draw[thick, blue] (s2)--++(216:1) node[black,below left = 0.1ex and 0.1ex] {$\omega_5^3$};
    \draw[thick, blue] (s2)--++(288:1) node[black,below right = 0.1ex and 0.1ex] {$\omega_5^4$};
  \end{tikzpicture}
  \caption{Example of a nontrivial sum of unit roots}
  \label{fig_nontrivial}
\end{figure}

The most important special case of a vanishing sum of unit roots is the sum
of all $m$-th unit roots, that is,
$\omega_m + \omega_m^2 + \cdots + \omega_m^m = 0$ holds for an integer
$m > 1$, which can be seen using a geometric sum. In the literature these
sums are often denoted as \emph{trivial}. An example of a nontrivial
vanishing sum of unit roots is
\begin{align}
  \omega_6 + \omega_6^5 + \omega_5 + \omega_5^2 + \omega_5^3 + \omega_5^4 = 0.
  \label{eq_nontrivialsum}
\end{align}
This sum can be written as the sum of
the trivial sum $\sum_{v=1}^5 \omega_5^v = 0$ and the rotated trivial sum
$\omega_6 + \omega_6^3 + \omega_6^5 = \omega_6^1\cdot (\sum_{v=1}^3 \omega_3^v) = 0$, see
Figure~\ref{fig_nontrivial}.

\begin{remark}\label{rem:NoEasyFormulation}
  An obvious way to find unit roots $\e^{j\sigma_\ell}$ and electrical
  angles $\Phi_i$ corresponding to an ambiguity of a linear array
  would be a feasibility problem with
  Constraints~\eqref{eq:ConnectionUnitrootsYT1}
  and~\eqref{eq:ConnectionUnitrootsYT2}. However, the
  real-valued reformulation of
  $\sum_\ell \e^{j\sigma_\ell} = 0$, with $\sigma_\ell \in [0,2\pi)$ is
  $\sum_\ell \cos(\sigma_\ell) = 0$ and $\sum_\ell \sin(\sigma_\ell) = 0$,
  which is a trigonometric constraint that is hard to handle in most
  solvers. The same problem emerges, if $\e^{j \sigma_\ell}$ is replaced by
  a complex variable $\omega_\ell$ for all~$\ell \in [M]$. This would
  yield~$\sum_{\ell} \Real(\omega_\ell) = \sum_{\ell} \Imag(\omega_\ell) =
  0$ and $\Real(\omega_\ell)^2 + \Imag(\omega_\ell)^2 = 1$. In this case,
  we need to control the argument of the complex variables $\omega_\ell$ to
  model Equation~\eqref{eq:ConnectionUnitrootsYT2}, which is nonlinear.
\end{remark}

In order to avoid the issue in Remark~\ref{rem:NoEasyFormulation}, we make
the following additional assumption.
\begin{enumerate}[resume*=assump]
\item\label{assump:SchurPoly} For every root of the Schur polynomial, the
  corresponding exponential sum ~$\sum_{\ell=1}^N \e^{j\sigma_\ell}$ is a
  vanishing sum of roots of unity, possibly rotated by a complex number on
  the unit circle.
\end{enumerate}
The subsequent Example~\ref{ex:LimitationOfApproach} demonstrates that
there exist linear arrays with ambiguities that violate this assumption.
However, the Assumption~\ref{assump:SchurPoly} will allow us to use a purely linear
procedure to find many ambiguities that are present in a linear array. This
procedure relies on the following theorem, which reduces the case of
general vanishing sums to minimal vanishing sums, i.e., vanishing sums, such
that no proper subsum also vanishes.



\begin{theorem}[Mann~\cite{Mann65}, Corollary 1.1]\label{thm:reducevansum1}
  Let $\sum_{i=1}^k a_i \, \eta_i = 0$ be a vanishing sum with
  $a_i \in \Z \setminus \{0\}$, $i \in [k]$, and unit roots $\eta_i$,
  $i \in [k]$. Then this sum can be written as
  \[
    \sum_{i=1}^k a_i \, \eta_i =\zeta_1 \, \sum_{i=1}^{k_1}a_i \, \nu_i +
    \dots + \zeta_u \, \sum_{i=k_{u-1}+1}^{k_u} a_i \, \nu_i,
  \]
  with unit roots $\zeta_1, \dots, \zeta_u$,
  $0 \enifed k_0 < k_1 < \cdots < k_u \define k$ and all $\nu_i$ are $(p_1 \, p_2 \, \cdots \, p_s)$-th
  unit roots with $0 < p_t \leq \max\{k_{j}-k_{j-1} \suchthat j \in [u] \}$, for prime numbers $p_t$, $t \in [s]$.
  Moreover, each vanishing sum
  $\sum_{i = k_j+1}^{k_{j+1}}a_i \, \nu_i = 0$, $j = 0,\dots,u-1$ is
  minimal.
\end{theorem}

Note that the upper bound 
for $p_t$ in Theorem~\ref{thm:reducevansum1} does not depend on $t$. Moreover,
the representation in
Theorem~\ref{thm:reducevansum1} is clearly not unique.\\

Because of Theorem~\ref{thm:reducevansum1}, we only need to consider minimal
vanishing sums, as they can be rotated and linearly combined to obtain
general rotated vanishing sums and thus ambiguities as roots of the Schur
polynomial. If we allow the linear coefficients~$\zeta_i$ in
Theorem~\ref{thm:reducevansum1} to be complex numbers on the unit circle,
i.e,~$\zeta_i = \e^{j2\pi v_i}$, $v_i \in [0,1)$, instead of roots of
unity, we are also able to find slightly more general exponential sums that sum
to zero and thus induce ambiguities.

\begin{remark}\label{rem:NoMinVanSum14}
  There are no (nontrivial) vanishing sums of
  length~1. Furthermore, every vanishing sum of length~4 can be written as
  the sum of two (rotated) minimal vanishing sums of length~2, and thus,
  there are no minimal vanishing sums of length~4, which can be deduced
  from~\cite[Lemma~3.2 and Lemma~3.3]{PoonenR98}.
\end{remark}


\begin{remark}[Limitation of the approach] At least for symmetric linear
arrays there exist ambiguities which cannot be represented as a linear
combination of minimal vanishing sums with rotation factors of the
form~$\e^{j2\pi v}$, $v \in [0,1)$ and thus violate Assumption~\ref{assump:SchurPoly}. In this case, the approach of this section is not able
to detect these ambiguities. The following example shows such an
ambiguity, which can be obtained by using~\cite[Theorem~7.1]{Manikas04}.
\end{remark}

\begin{example}\label{ex:LimitationOfApproach}
  Consider the linear array $\br = (0,1,3,4)\T$ with baseline~$d=1$ and
  $\blambda = (0,0,1,1)\T$. The Schur polynomial for~$\br$ is given by
  \begin{align*}
    s_\blambda =\ & \e^{j(\Phi_1+\Phi_2)} + \e^{j(\Phi_1+\Phi_3)} +
                  \e^{j(\Phi_1+\Phi_4)}\\
    & + \e^{j(\Phi_2+\Phi_3)} + \e^{j(\Phi_2+\Phi_4)} +
    \e^{j(\Phi_3+\Phi_4)}.
  \end{align*}
  The ambiguity
  \begin{align*}
      \bPhi & = \big[ 2\arctan(\Delta_1) - \pi,\, 2\arctan(\Delta_2)-\pi,\,\\
      & \qquad \pi-2\arctan(\Delta_2),\, \pi-2\arctan(\Delta_1) \big]\T\\
    & \approx (43.6\degree,\, 77.35\degree,\, 102.7\degree,\,
    136.4\degree)\T,
  \end{align*}  
  \begin{align*}
    \text{with }\Delta_1 = \sqrt{\tfrac{1}{3}\big(12 - \sqrt{129}\big)},\;
    \Delta_2 = \sqrt{\tfrac{1}{3}\big(12 + \sqrt{129}\big)},
  \end{align*}
  yields $\Phi_1 + \Phi_4 = \Phi_2 + \Phi_3$, whereas all other exponents
  of~$s_\blambda$ are pairwise different. This means, in order to encode
  the solution $\bPhi$ by summing rotated minimal vanishing sums of roots
  of unity, two of the six resulting roots of unity must be equal (after a
  possible rotation). Since there are $N=6$ SSYTs for the linear
  array~$\br$, there are three ways of summing rotated minimal vanishing
  sums in order to obtain a vanishing sum of length~6, namely,
  $6 = 3 + 3 = 2 + 2 + 2$. Note that there are no minimal vanishing sums of
  lengths~$1$ and~$4$, see Remark~\ref{rem:NoMinVanSum14}.  However, it is
  easy to see that for all three cases, it is not possible to find rotation
  factors such that two roots of unity are equal and five are pairwise
  different. Thus, the ambiguity given by~$\bPhi$ cannot be represented
  using sums of rotated minimal vanishing sums, so that
  Assumption~\ref{assump:SchurPoly} is violated.
\end{example}

\section{Enumeration Using Minimal Vanishing Sums}
\label{sec:EnumMVS}

\noindent
Recall that finding ambiguous DoAs for the linear array
$\br = \blambda + \bdelta$ corresponds to finding roots of the Schur
polynomial $s_{\blambda}$ in~\eqref{eq:schurpolyrepresentation}. The Schur
polynomial is given by a sum of exponentials, where the exponents need to
fulfill a linear relation modulo~$2\pi$, which originates from the SSYTs of
shape~$\blambda$ for the linear array~$\br$,
see~\eqref{eq:ConnectionUnitrootsYT1}
and~\eqref{eq:ConnectionUnitrootsYT2}. We assumed that~$s_\blambda$ can be
represented using linear combinations of minimal vanishing sums with
coefficients~$\zeta = \e^{j2\pi v}$, $v \in [0,1)$. By that, we can find a
fairly large subset of roots of the Schur polynomial. If~$N$ is the number
of SSYTs of shape~$\blambda$, then the following procedure can be used to
find ambiguities in~$\br$.

\begin{enumerate}[leftmargin=3ex]
\item Choose a partition~$N = p_1 + \cdots + p_k$, $k \in [N]$,
  with $p_i \in [N]\setminus \{1,4\}$, i.e., without using the elements~$1$ and~$4$,
  since there are no minimal vanishing sums of unit roots of these lengths, see
  Remark~\ref{rem:NoMinVanSum14}.
\item For each partition element~$p_i$ choose a minimal vanishing sum $S_i$ of
  length~$p_i$.
\item Choose an assignment of the roots of unity appearing in~$S_i$ to the
  variables $\sigma_\ell$, $\ell \in [N]$.
\item Check if there exists a solution to the linear equation
  system~\eqref{eq:ConnectionUnitrootsYT2}, where each $S_i$ can be rotated
  by an arbitrary $\e^{j2\pi v_i}$, with~$v_i \in [0,1)$,
  c.f. Theorem~\ref{thm:reducevansum1} and the discussion thereafter.
\end{enumerate}
Clearly, for this approach to work, we need to know all possible minimal
vanishing sums that can be used to build the desired vanishing sum. For
$N \leq 12$, all minimal vanishing sums (up to rotations) of length $N$ are
characterized by Poonen and Rubinstein~\cite[Table~1 and
Theorem~3]{PoonenR98}\footnote{Mann~\cite{Mann65} already characterized all
  minimal vanishing sums up to $N = 7$, and Conway and
  Jones~\cite{ConwayJ76} characterized all minimal vanishing sums up to
  $N = 9$.}, and their construction can be easily extended to $N > 12$.

\subsection{The MIP-Formulation}\label{sec:MIPFormulation}

\noindent
The approach described above to find ambiguities for a given linear
array~$\br$ can be formulated as a mixed-integer (linear) program
(MIP). Every feasible solution of the MIP then corresponds to an ambiguity
for~$\br$, such that by enumerating feasible solutions of the MIP, it is
possible to obtain many ambiguities for~$\br$. The only requirement is that
all minimal vanishing sums with length up to~$N$ need to be known in
advance. The following parameters are used in the
MIP-formulation~\eqref{eq:FeasibilityProblem} in Figure~\ref{fig:MIP}:
\begin{align*}
  M  \define\; & \text{Number of sensors in the linear array}\\
               & \br = \blambda + \bdelta, \text{ i.e.}, \br \in \Z^M, \\
    d  \define\; & \text{common baseline of the array positions}~\br,\\
    N  \define\; & \text{number of SSYTs of shape $\blambda$}, \\
    \bP  \define\; & [\underbrace{2,\dots,2}_{\lfloor \tfrac{N}{2} \rfloor \text{ many}},
                 \underbrace{3,\dots,3}_{\lfloor \tfrac{N}{3} \rfloor \text{ many}},
                 \underbrace{5,\dots,5}_{\lfloor \tfrac{N}{5} \rfloor \text{ many}},
                 \underbrace{6,\dots,6}_{\lfloor \tfrac{N}{6} \rfloor \text{ many}},
                 \dots,n], \\
    m_i \define\;& \text{\# minimal vanishing sums of length $P_i$}\\
  u^{(i)}_{t,k} \define\;& \text{$k$-th root of unity of $t$-th minimal vanishing}\\
                  & \text{sum of length $P_i$}, \; i \in [\abs{P}], \; t \in [m_i], \, k \in [P_i], \\
    I \define\;& \{(i,t,k) \suchthat i \in [\abs{\bP}], \; t
               \in [m_i],\, k \in [P_i]\}.
\end{align*}
The vector $\bP$ is needed to model all  partitions of $N$ that can be used in the
first step of the procedure above, where~$\abs{\bP}$ denotes its
number of entries.

The variables are:
\begin{itemize}[leftmargin=3ex]
  \item $\bq^{(i)} \in \{0,1\}^{m_i}$ with $q^{(i)}_t = 1$ if and only if
    $P_i$ is part of the chosen partition of~$N$ and the $t$-th minimal
    vanishing sum is selected for this $P_i$.
  \item $\bb^{(i)} \in \{0,1\}^{m_i \times P_i \times N}$. We have
    $b^{(i)}_{t,k,\ell} = 1$ if and only if $\sigma_\ell$ is assigned to the $k$-th
    root of unity in the $t$-th minimal vanishing sum chosen for the $i$-th
    element of the partition.
  \item $\bv^{(i)} \in [0,2\pi)^{m_i}$: rotation factors for the minimal
    vanishing sums of length~$P_i$.
  \item $\bw^{(i)} \in [0,2\pi)^{m_i \times P_i \times N}$: auxiliary variables
    for linearization:
    $w^{(i)}_{t,k,\ell} = v^{(i)}_t\cdot b^{(i)}_{t,k,\ell}$.
  \item $\bz \in \Z^N$: models that a rotation is always applied
    modulo~$2\pi$. Since the rotation values lie in $[0,2\pi)$, all $z_\ell$
    can be assumed to be binary.
  \item $\bsigma \in [0,2\pi)^N$, $\bPhi \in [-\pi d,\pi d]^M$, $\bx \in \Z^N$:
    model Equations~\eqref{eq:ConnectionUnitrootsYT1}
    and~\eqref{eq:ConnectionUnitrootsYT2}, where $x$ is needed for the
    modulo operation. Constraint~\eqref{eq:YoungTableaux} implies that each $x_\ell$ is bounded by
    $-\frac{1}{2}d\sum_{m=1}^M \alpha_{m,\ell} - 1 \leq x_\ell
    \leq \frac{1}{2}d(\sum_{m=2}^M \alpha_{m,\ell} - \alpha_{1,\ell})$, for
    $\ell \in [N]$, see Constraint~\eqref{eq:Boundx}.
  \end{itemize}


\begin{figure*}[tb]\centering  
  \begin{subequations}
    \begin{align}
      \sum_{i=1}^{\abs{P\,}}P_i\cdot\Big(\sum_{t=1}^{m_i}q^{(i)}_t\Big)
      &= N,
      && \label{eq:Partition1} \\
      \sum_{t=1}^{m_i} q^{(i)}_t
      & \leq 1
      &&\forall\, i \in [\abs{P}], \label{eq:Partition2} \\
      b^{(i)}_{t,k,\ell}
      &\leq q^{(i)}_t
      && \forall\, (i,t,k,\ell)\in I\times [N], \label{eq:Couplebq} \\
      \sum_{i=1}^{\abs{P\,}} \sum_{t=1}^{m_i} \sum_{k=1}^{P_i}
      b^{(i)}_{t,k,\ell} &= 1,
      && \forall\, \ell \in [N], \label{eq:ChooseRoots1} \\
      \sum_{\ell=1}^N b^{(i)}_{t,k,\ell}
      &\leq 1,
      && \forall\, (i,t,k) \in I \label{eq:ChooseRoots2}, \\ 
      \sum_{i=1}^{\abs{P\,}} \sum_{t=1}^{m_i} \sum_{k=1}^{P_i}
      w^{(i)}_{t,k,\ell}  +b^{(i)}_{t,k,\ell}\, u^{(i)}_{t,k}  - 2\pi\,z_\ell
      &=\sigma_{\ell},
      && \forall\, \ell \in [N],\label{eq:AssignRoots1} \\
      w^{(i)}_{t,k,\ell} &\leq b^{(i)}_{t,k,\ell} \cdot 2\pi,
      && \forall\,
      (i,t,k,\ell) \in I \times [N], \label{eq:AssignRoots2} \\
      2\pi(-1 + b^{(i)}_{t,k,\ell}) + w^{(i)}_{t,k,\ell} \leq v_t^{(i)}
      &\leq 2\pi(1 - b_{t,k,\ell}^{(i)}) + w^{(i)}_{t,k,\ell},
      &&\forall\, (i,t,k,\ell) \in I \times [N], \label{eq:AssignRoots3} \\
      \sum_{m=1}^M \alpha_{m,\ell}\, \Phi_m - 2\pi\, x_\ell  &= \sigma_\ell,
      && \forall\, \ell \in [N], \label{eq:YoungTableaux} \\
      -\pi d = \Phi_1 < \Phi_2 < \cdots < \Phi_M
      &\leq \pi d, \label{eq:UniquePhi} \\
      -\frac{1}{2}d\sum_{m=1}^M \alpha_{m,\ell} - 1 \leq x_\ell
      &\leq \frac{1}{2}d\big(\sum_{m=2}^M \alpha_{m,\ell} - \alpha_{1,\ell}\big) 
      && \forall\, \ell \in [N], \label{eq:Boundx} \\
      \omit\rlap{$\displaystyle \bq^{(i)} \in \{0,1\}^{m_i},\; \bb^{(i)} \in
      \{0,1\}^{m_i \times P_i \times N} ,\, \bz \in \{0,1\}^N,\; \bx \in \Z^N,\;
      \bw^{(i)} \in [0,2\pi)^{m_i \times P_i \times N}  $} \notag \\
      \omit\rlap{$\displaystyle \bv^{(i)} \in
      [0,2\pi)^{m_i},\; \bPhi\in [-\pi d, \pi d]^M,\; \bsigma \in [0,2\pi)^N.$} \notag
    \end{align}
    \label{eq:FeasibilityProblem}
  \end{subequations}
  \caption{Feasibility MIP for enumerating all ambiguities.}
  \label{fig:MIP}
\end{figure*}

Constraints~\eqref{eq:Partition1} and~\eqref{eq:Partition2} ensure that a
valid partition of~$N$ together with corresponding minimal vanishing sums
is chosen.  The subsequent Constraints~\eqref{eq:Couplebq}--\eqref{eq:AssignRoots3} model the assignment of the chosen roots of
unity to the variables~$\sigma_\ell$, using a linearization of the bilinear
constraint 
\begin{align*}
  \sum_{i=1}^{\abs{P\,}} \sum_{t=1}^{m_i} \sum_{k=1}^{P_i}
      b^{(i)}_{t,k,\ell}\, (v^{(i)}_t + u^{(i)}_{t,k})  - 2\pi\,z_\ell
      =\sigma_{\ell}, \; \forall\, \ell \in [N].
\end{align*}
Finally, Constraint~\eqref{eq:YoungTableaux} models the linear
equations~\eqref{eq:ConnectionUnitrootsYT2}.  Because of
Assumption~\ref{assump:FixFirstDoA}, we can fix the first
electrical angle $\Phi_1 = -\pi d$ in
Constraint~\eqref{eq:UniquePhi}. In terms of vanishing sums, this amounts
to a global rotation of all appearing roots of unity which does not destroy
the sum being $0$, as can be seen by
\begin{align}\label{eq:FixingPhi}
  \sum_{\ell=1}^N \e^{j(2\pi v + \mu)} =
  \sum_{\ell=1}^N \e^{j2\pi v} \cdot \e^{j\mu}.
\end{align}
In order to remove some symmetric solutions, the variables~$\bPhi$ are be
ordered increasingly in Constraint~\eqref{eq:UniquePhi}. This is justified
by Theorem~\ref{thm:OrderingFixing} in the next section. Additionally, the strict inequalities prevent trivial
ambiguities consisting of two or more equal electrical angles. These strict inequalities
together with the upper bounds for $\bv^{(i)}$, $\bw^{(i)}$, $\bPhi$ and $\bsigma$ are
modeled using non-strict inequalities with a small $\epsilon = 0.001$, so
that~\eqref{eq:FeasibilityProblem} is indeed a MIP.

Note that there could exist different feasible solutions that correspond to
the same ambiguous vector of electrical angles. There are different reasons for this
behavior. First, the decomposition of a vanishing sum of unit roots into
minimal vanishing sums is not unique, such that different sums of (rotated)
minimal vanishing sums of unit roots can lead to the same vanishing sum and
thus, to the same ambiguity. Second, there can be different assignments of
the chosen unit roots to the variables~$\sigma_\ell$ that lead to the same
electrical angles~$\Phi_m$.

\subsection{Analysis of the MIP formulation}

\begin{theorem}\label{thm:OrderingFixing}
  Consider a feasible solution 
  $\bX = (\bb^{(i)}, \, \bq^{(i)}, \, \bx,\, \bz,\, \bv^{(i)}, \, \bw^{(i)}, \, \bPhi,\,
  \bsigma)$ for~\eqref{eq:FeasibilityProblem} and
  let $\tau \in S_M$ be an arbitrary permutation of $[M]$. Then there exists a feasible solution
  $\tilde{\bX} = (\tilde{\bb}^{(i)},\, \tilde{\bq}^{(i)},\, \tilde{\bx},\,
  \tilde{\bz},\, \tilde{\bv}^{(i)},\, \tilde{\bw}^{(i)},\, \tilde{\bPhi},\,
  \tilde{\bsigma})$ with
  $\tilde{\bPhi} = (\Phi_{\tau(1)},\dots,\Phi_{\tau(M)}) \in [-\pi d,\pi d]^M$.
\end{theorem}

Note that the statement is not trivial: If the generalized Vandermonde determinant is 0, permuting~$\bPhi$ does not
change this, and it does not change the Schur polynomial. However, it is not
clear whether there still exists a sum of rotated minimal vanishing sums and
assignment of the appearing roots of unity so that
Constraint~\eqref{eq:YoungTableaux} is satisfied.

\begin{proof}
  Let $\br = \blambda + \bdelta$ be a linear array, and let
  $\mathcal{T}_{\blambda}$ be the set of all SSYTs of shape $\blambda$.
  We first prove the assertion that if there exists an SSYT
  $T_{\ell} \in \mathcal{T}_{\blambda}$ with weight vector
  $\balpha(T_{\ell}) = (\alpha_{1,\ell},\dots,\alpha_{M,\ell})$, then for all
  permutations $\tau \in S_M$ there also exists an SSYT
  $T_{k} \in \mathcal{T}_{\blambda}$ with
  $\balpha(T_{k}) = (\alpha_{1,k},\dots,\alpha_{M,k}) =
  (\alpha_{\tau^{-1}(1),\ell},\dots,\alpha_{\tau^{-1}(M),\ell}) \enifed
  \tau(\balpha(T_\ell))$.

  Let~$\bbeta$, $\bbeta'$ be two possible weight vectors of an SSYT of
  shape~$\blambda$ so that~$\bbeta$ and~$\bbeta'$ only differ by swapping two
  consecutive entries. Then there exists a bijection between the
  SSYTs of shape~$\blambda$ with weight vector~$\bbeta$ and~$\bbeta'$:
  Consider the entries~$\beta_i$ and~$\beta_{i+1}$ of the weight
  vector~$\bbeta$, and let~$T$ be an SSYT of shape~$\blambda$ with weight
  vector~$\bbeta$. Select all columns of~$T$ that contain exactly one entry
  equal to~$i$ or~$i+1$. All other columns contain either no or two such
  entries. In each row of~$T$ replace each~$i$ appearing in these
  columns by~$i+1$ and vice versa. After reordering
  the rows so that these are again sorted nondecreasingly, we obtain an
  SSYT of shape~$\blambda$ with weight vector~$\bbeta'$. This yields the
  desired bijection, see also~\cite[Proof of Theorem~7.10.2]{Stanley97}.
  Thus, for a given weight vector~$\balpha$ and a given
  permutation~$\tau \in S_M$ we can find an SSYT with weight
  vector~$\tau(\balpha)$ by decomposing~$\tau$ into a sequence of transpositions of
  consecutive entries and using the compositions of the respective
  bijections.
  

  Consider the solution $\bX$ and permutation $\tau$. In order to prove the
  existence of solution $\tilde{\bX}$ with $\tilde{\Phi}_m = \Phi_{\tau(m)}$, $m \in [M]$, we show that
  there exists a permutation $\gamma \in S_N$ with
  $\tilde{\sigma}_\ell = \sigma_{\gamma(\ell)}$:
  \begin{align*}
    \tilde{\sigma}_{\ell}
    &= \alpha_{1,\ell} \, \tilde{\Phi}_1 + \dots + \alpha_{M,\ell} \,
      \tilde{\Phi}_M  - 2\, \tilde{x}_{\ell} \\
    &= \alpha_{1,\ell} \, \Phi_{\tau(1)} + \dots + \alpha_{M,\ell} \,
      \Phi_{\tau(M)} - 2\, \tilde{x}_{\ell} \\
    &= \alpha_{\tau^{-1}(1),\ell} \, \Phi_{1} + \dots + \alpha_{\tau^{-1}(M),\ell} \,
      \Phi_{M} - 2\, \tilde{x}_{\ell}.
  \end{align*}
  By the assertion, there exists an SSYT $T$ with weight vector
  $\tau(\balpha(T_\ell))$.
  Defining the permutation~$\gamma$ such that
  $\alpha_{m,\gamma(\ell)} = \alpha_{\tau^{-1}(m),\ell}$ for all $m \in [M]$, i.e., such that the
  SSYT~$T_\gamma(\ell)$ has weight vector~$\tau(\balpha(T_\ell))$, and setting
  \begin{align*}
    &\tilde{\sigma}_\ell = \sigma_{\gamma(\ell)}, \; \tilde{x}_\ell =
    x_{\gamma(\ell)}, \; \tilde{z}_\ell = z_{\gamma(\ell)}, \;
    \tilde{b}^{(i)}_{t,k,\ell} = b^{(i)}_{t,k,\gamma(\ell)},\\
    & \tilde{w}^{(i)}_{t,k,\ell} = w^{(i)}_{t,k,\gamma(\ell)}, \;
    \tilde{q}^{(i)}_t = q^{(i)}_t, \; \tilde{v}^{(i)}_t = v^{(i)}_t,
  \end{align*}
  for all $(i,t,k,\ell) \in I \times [N]$ yields the desired feasible
  solution $\tilde{\bX}$ with $\tilde{\bPhi} \in [-\pi d,\pi d]^M$ and
  $\tilde{\Phi}_m = \Phi_{\tau(m)}$ for all $m \in [M]$.
\end{proof}

The following two Lemmas state that for a linear array with
integer positions, the feasibility
problem~\eqref{eq:FeasibilityProblem} finds all ambiguities in
the array
that can be represented using a linear combination of minimal vanishing
sums with coefficients on the complex unit circle.

\begin{lemma}\label{lem:FeasSoluMIPisDoA}
  For a linear array with positions corresponding to integer multiples
  $\br \in \Z^M$ of a common baseline~$d \leq 1$ measured in half
  wavelength, each feasible solution of the
  feasibility-MIP~\eqref{eq:FeasibilityProblem} corresponds to an ambiguous
  vector of electrical angles.
\end{lemma}

\begin{proof}
  Let
  $\bX = (\bb^{(i)}, \, \bq^{(i)}, \, \bx,\, \bz,\, \bv^{(i)}, \, \bw^{(i)}, \, \bPhi,\,
  \bsigma)$ be a feasible solution for~\eqref{eq:FeasibilityProblem}. It is
  clear by construction that the Schur polynomial~$s_\blambda(\bz)$
  with~$z_i = \e^{j \Phi_i}$ satisfies $s_\blambda(\bz) = 0$. By definition,
  this implies that the generalized Vandermonde determinant~$V_\br(\bz)$
  vanishes. Thus, the array steering matrix $\bA(\bPhi)$ is rank-deficient, i.e.,
  $\Phi_1,\dots,\Phi_M$ are ambiguous.
\end{proof}

The next Lemma is an immediate consequence of the definition of the Schur
polynomial and the arguments above.
\begin{lemma}\label{lem:DoAisFeasSoluMIP}
  Let $\br \in \Z^M$ be an arbitrary integer linear array where the
  positions are multiples of a common baseline~$d \leq 1$ measured in half wavelength. Each
  ambiguous vector of electrical angles that forms a root of the Schur polynomial and
  that can be represented using sums of rotated minimal vanishing sums,
  corresponds to at least one feasible solution of the
  feasibility-MIP~\eqref{eq:FeasibilityProblem}.
\end{lemma}

Let us now relate our approach to the \emph{uniform ambiguities}
from~\cite[Theorem~2.2]{ManikasP98}. For the ease of presentation, we
restate this result in terms of electrical angles.

\begin{theorem}[\cite{ManikasP98}]\label{thm:uniformpartamb}
  Let $\br = (r_1,\dots,r_M)\T \in \R^M$ be an arbitrary linear array with
  positions measured in half wavelength and baseline~$d=1$. Define the
  vector $\bPhi_{i,j}$ of electrical angles as
  \begin{align*}
    \bPhi_{i,j} \define \!
    \bigg[ & \!\! -\pi,\, -\pi\Big(1-\frac{2}{\abs{r_i-r_j}}\Big),\,
    -\pi\Big(1-\frac{4}{\abs{r_i-r_j}}\Big),\\
    & \dots,\; -\pi\Big(1-\frac{2c}{\abs{r_i-r_j}}\Big)\bigg]\T,
  \end{align*}
  where $i\neq j \in [M]$ and $c \in \N$ is the largest integer satisfying
  $c < \abs{r_i-r_j}$. Then, if $\bPhi_{i,j}$
  contains at least $M$ elements, any subvector of $M$ elements from
  $\bPhi_{i,j}$ 
  is an ambiguous vector.
\end{theorem}

Note that these are all ambiguities that are currently known in the
literature for non-symmetric integer linear arrays.

We can now prove our main result, namely, that for an integer linear array,
our approach is able to identify all \emph{uniform ambiguities}
from~Theorem~\ref{thm:uniformpartamb}. This directly implies we can find all
all ambiguities previously known in the literature for \emph{non-symmetric} integer linear arrays.

\begin{proposition}\label{prop:FindAllUniformAGS}
  Let $\br \in \Z^M$ be an arbitrary integer linear array where the
  positions are multiples of a common baseline~$d \leq 1$ measured in half
  wavelength. Let~$\tilde{\br} = d\cdot \br \in \R^M$, be the array~$\br$
  rescaled to a baseline~$\tilde{d} = 1$, such that~$\br$ and~$\tilde{\br}$
  are in fact two representations (with different baselines) of the same
  linear array. Then any ambiguity of the form $\bPhi_{i,j}$ as stated in
  Theorem~\ref{thm:uniformpartamb} (for the representation~$\tilde{\br}$)
  corresponds to at least one solution of the
  feasibility-MIP~\eqref{eq:FeasibilityProblem} (for the representation~$\br$).
\end{proposition}

\begin{proof}
  Let $\bPhi \in \R^M$ be an ambiguity for~$\tilde{\br}$ in the form of
  Theorem~\ref{thm:uniformpartamb}, and assume w.l.o.g. that the first
  electrical angle is~$-\pi d$, i.e., $\Phi_1 = -\pi d$ and for
  $m = \{2, \dots, M\}$:
  \begin{align*}
    \Phi_m =
    -\pi d\Big(1-\frac{2c_m}{\abs{\tilde{r}_i-\tilde{r}_j}}\Big)
    = -\pi d\Big(1-\frac{2c_m}{d\,\abs{r_i-r_j}}\Big),
  \end{align*}
  with $i\neq j \in [M]$, $c_m \in \N$ and $c_{m_1} \neq c_{m_2}$ for all
  $m_1 \neq m_2 \in [M]$.
  The variables $\sigma_\ell$ in~\eqref{eq:FeasibilityProblem} are then given by
  \begin{align*}
    \sigma_\ell = \sum_{m=1}^M \alpha_{m,\ell}\Phi_m \; \mod \; 2\pi,
  \end{align*}
  for $m \in [M]$ and $\ell \in [N]$. Thus,
  \begin{align}
    & \sum_{\ell=1}^N e^{j\sigma_\ell}
    = \sum_{\ell=1}^N \exp\Big(j\sum_{m=1}^M \alpha_{m,\ell}(\Phi_m + \pi
      d - \pi d)\Big) \notag \\
    &= e^{-j\pi d K} \sum_{\ell=1}^N \exp\Big( j \sum_{m=1}^M
      \alpha_{m,\ell}(\Phi_m + \pi d)\Big), \label{eq:PMAmbVanSum}
  \end{align}
  where~$K = \sum_{m=1}^M \alpha_{m,\ell}$ is the same constant for
  all~$\ell \in [N]$. 
  Since $\Phi_m +\pi d \in \{2\pi\,w
  \suchthat w \in \Q\}$ for all $m \in [M]$ it holds
  that~\eqref{eq:PMAmbVanSum} is a sum of roots of unity, rotated by a
  complex number on the complex unit circle.
  
  Since $\bPhi$ forms an ambiguous vector of
  electrical angles for the linear array $\br$, $z_m =
  \exp(j \Phi_m)$ form a root of the generalized Vandermonde
  determinant $V_{\br}(\bz)$ and thus of the Schur polynomial $s_{\blambda}(\bz)$, where
  $\blambda = \br -
  (0,1,\dots,M-1)$. Equations~\eqref{eq:schurpolyrepresentation},~\eqref{eq:ConnectionUnitrootsYT1}
  and~\eqref{eq:ConnectionUnitrootsYT2} imply $0 = s_{\blambda}(\bz) = \sum_{\ell=1}^N \exp(j\,\sigma_\ell)$,
  and thus~\eqref{eq:PMAmbVanSum} is a rotated vanishing sum of roots of
  unity. Theorem~\ref{thm:reducevansum1} implies
  that~\eqref{eq:PMAmbVanSum} can be written as linear combination of
  minimal vanishing sums with coefficients on the complex unit circle. By
  Lemma~\ref{lem:DoAisFeasSoluMIP}, there is at least one feasible solution
  of the MIP~\eqref{eq:FeasibilityProblem} that corresponds to the
  ambiguity $\bPhi$, which finishes the proof.
\end{proof}

Moreover, the computational results in Section~\ref{sec:CompRes} below show
that for non-symmetric integer linear arrays, our approach finds many more
ambiguities than previously known.
In general, it remains an open question, whether there also exist ambiguities
for non-symmetric integer linear arrays which cannot be expressed as linear
combination of minimal vanishing sums with coefficients on the complex unit
circle, or if our approach indeed finds \emph{all} ambiguities that are
present in such an array.

Before presenting computational results, let us shortly discuss some
details of enumerating the feasible solutions with our approach in the
next section.

\subsection{Enumerating all Feasible Solutions of the MIP}
\label{sec:EnumeratingMIP}

\noindent
Since it is possible that the MIP~\eqref{eq:FeasibilityProblem} has
infinitely many feasible solutions, we do not count all feasible solutions,
but only all configurations of the integer and binary variables
$\bb,\, \bq,\, \bx,\, \bz$ such that after fixing all these variables the
remaining problem has at least one feasible solution. Because all integer
and binary variables are bounded, there are only finitely many
configurations that need to be checked.

After all feasible configurations of the integer and binary variables have
been found, we apply a post-processing step in order to find the
ambiguities corresponding to each configuration. 
Due to the rotations $v^{(i)}_t$ of the used minimal vanishing sums, there
can exist configurations of the integer (and binary) variables with
infinitely many feasible solutions for the continuous
variables. These solutions form a whole class of ambiguities with a number
of parameters, see Example~\ref{ex:InfiniteSolutions}. Altogether, we end up with either finitely many ambiguities,
or finitely many classes of ambiguities, each of them depending on a number
of parameters. The ambiguities can be converted into DoAs by using
$-\pi d\cos(\theta_m) = \Phi_m$.

\begin{remark}\label{rem:DivideIntoPartitions}~
  In order to reduce the computational effort of enumerating all feasible
  integer solutions of~\eqref{eq:FeasibilityProblem}, the problem can
  be divided into smaller subproblems, one for each possible
  partition. In each of the smaller MIPs, most of the variables
  $\bq^{(i)} \in \{0,1\}^{m_i}$ can be fixed, according to the corresponding
  partition. Moreover, this eliminates many symmetric solutions in terms of
  the variables $q^{(i)}_j$.

\end{remark}

\section{Computational Results}
\label{sec:CompRes}

\noindent
In this section, we use the approach of obtaining ambiguities by
enumerating the feasible solutions of~\eqref{eq:FeasibilityProblem}
to identify ambiguities for some exemplary integer linear arrays. To
enumerate all feasible solutions we use the counting feature of
SCIP~6.0.0~\cite{SCIP600}. We use CPLEX~12.7.1.0 as LP solver. 
The first Example~\ref{ex:InfiniteSolutions} was
enumerated using the full Problem~\eqref{eq:FeasibilityProblem}, whereas for
Example~\ref{ex:LargeArray}, we divided Problem~\eqref{eq:FeasibilityProblem} into smaller ones, one for each possible
partition 
as described
in Remark~\ref{rem:DivideIntoPartitions}. In all examples, the baseline
is~$d=1$. The used model files can be obtained
via the website of the last author.

The computation for Example~\ref{ex:InfiniteSolutions}
was performed on a Linux desktop with 3.6~GHz Intel
Core i7-7700 Quad-Core CPUs having 16~GB main memory and 8~MB cache,
whereas the computations for Example~\ref{ex:LargeArray} were done on on a Linux cluster with 3.5 GHz Intel
Xeon E5-1620 Quad-Core CPUs, having 32~GB main memory and 10~MB cache. All
computations were performed single-threaded and, in the case of
Example~\ref{ex:LargeArray} with a timelimit of 450\,000\,s. For this
Example, we also display the time needed to enumerate all feasible solutions,
the total number of solutions that were enumerated, as well as the number
of processed nodes in the enumeration process.

Example~\ref{ex:InfiniteSolutions} shows a linear array with an infinite
number of ambiguities, even after fixing the first electrical angle to~$-\pi d$.

\begin{example}\label{ex:InfiniteSolutions}
   Consider the linear array $\br = \blambda + \bdelta = (0, 1, 3, 4)\T$, where
   $\bdelta = (0, 1, 2, 3)\T$ and $\blambda = (0, 0, 1, 1)\T$. The possible
   Young tableaux are shown in Figure~\ref{fig_tableaux_ex2}.
   
   \begin{figure}[h!]
     \centering
     \begin{ytableau}
       1 \\
       2
     \end{ytableau}
     \quad\ 
     \begin{ytableau}
       1 \\
       3
     \end{ytableau}
     \quad\ 
     \begin{ytableau}
       1 \\
       4
     \end{ytableau}
     \quad\
     \begin{ytableau}
       2 \\
       3
     \end{ytableau}
     \quad\ 
     \begin{ytableau}
       2 \\
       4
     \end{ytableau}
     \quad\ 
     \begin{ytableau}
       3 \\
       4
     \end{ytableau}\\
     \caption{Young tableaux for $\blambda = (0, 0, 1, 1)\T$}
     \label{fig_tableaux_ex2}
   \end{figure}
  
  \noindent
  This results in the Schur polynomial
  \begin{align*}
    s_{\blambda}(\bz) = z_1 z_2 + z_1 z_3 + z_1 z_4 + z_2 z_3 + z_2 z_4 + z_3 z_4.
  \end{align*}  
  Since $\br$ has $N = 6$ corresponding SSYTs, there are three possible
  partitions: $6 = 2+2+2 = 3+3 = 6$.\footnote{Note that there are no
    minimal vanishing sums of length 1 or 4, see
    Remark~\ref{rem:NoMinVanSum14}. 
  } For the partitions $6 = 2 + 2 + 2$ and $6 = 3 + 3$, there exist
  infinitely many valid solutions for the feasibility
  problem~\eqref{eq:FeasibilityProblem}, and thus infinitely many
  ambiguities, which can be classified into three classes.

  The partition $6 = 2 + 2 + 2$ yields the two infinite classes of electrical angles
  \begin{align*}
    -\pi \cdot \big[1,\, v,\, 0,\, -v\big]\T, \quad
    -\pi \cdot \big[1,\, 1-v,\, 1-2v,\, -v \big]\T,
  \end{align*}
  for all $v \in (0,1)$. The partition $6 = 3 + 3$ yields the following class of electrical angles 
  for $v \in (-1,\, 1)$ and $\gamma = \arccos(-\pi v)$:
  \begin{align*}
    -\pi \cdot \big[1,\,\tfrac{1}{3},\, -\tfrac{1}{3},
    v\big]\T \; \approx \;
    [0\degree,\, 70.53\degree,\, 109.47\degree,\, \gamma]\T.
  \end{align*}
  Since already the steering vectors corresponding to the electrical angles
  $-\pi \cdot [1,\,\tfrac{1}{3},\, -\tfrac{1}{3}]\T$ are linearly dependent and
  thus induce a rank-deficient steering matrix, an arbitrary electrical angle
  $v \in (-\pi,\,\pi)$ can be added to the three electrical angles in
  order to obtain an ambiguity with four electrical angles.

  For the third partition $6 = 6$, there are finitely many corresponding
  solutions, namely the eight ambiguities in
  Table~\ref{tab:r0134UniqueAmbs}.  Note that all three ambiguities that
  were already found by using the methods from Proukakis and
  Manikas~\cite{ManikasP98}, are contained in
  one of the three classes.
  
  Some of the non-uniform ambiguities obtained with the methods for
  symmetric linear arrays in~\cite{Manikas04} are contained in
  $-\pi\cdot [1,\, v,\, 0,\, -v]\T$, but there
  also exist non-uniform ambiguities which cannot be found with our
  approach, see Example~\ref{ex:LimitationOfApproach}.

  Altogether, our approach finds two new classes of ambiguities for this
  particular linear array that have not been known in the literature. Additionally, all previously known ambiguities
  which can be expressed as sums of rotated minimal vanishing sums are
  found by our approach as well.
  
  \begin{table}[tb]
    \caption{Unique ambiguities for Partition $6 = 6$ in the linear array
    $\br = (0,1,3,4)$.}
    \label{tab:r0134UniqueAmbs}
    \centering
    {\footnotesize
      \begin{tabular*}{\linewidth}{@{}l@{\extracolsep{\fill}}l@{}}
        \toprule Electrical Angle & Degree \\
        \midrule
        $-\pi\, [1,\, 14/15, 8/15,\, -3/15]$
                         & $[0\degree,\, 21.04\degree,\, 57.77\degree,\, 101.54\degree]$ \\
        $-\pi\, [1,\, 14/15, 2/15,\, -9/15] $
                         & $[0\degree,\, 21.04\degree,\, 82.34\degree,\, 126.87\degree]$ \\     
        $-\pi\, [1,\, 9/15, 8/15,\, -4/15] $
                         & $[0\degree,\, 53.13\degree,\, 57.77\degree,\, 105.47\degree]$ \\
        $-\pi\, [1,\, 9/15, -2/15,\, -14/15] $
                         & $[0\degree,\, 53.13\degree,\, 97.66\degree,\, 158.96\degree]$ \\
        $-\pi\, [1,\, 4/15, -2/15,\, -3/15] $
                         & $[0\degree,\, 74.53\degree,\, 97.66\degree,\, 101.54\degree]$ \\
        $-\pi\, [1,\, 4/15, -8/15,\, -9/15] $
                         & $[0\degree,\, 74.53\degree,\, 122.23\degree,\, 126.87\degree]$ \\
        $-\pi\, [1,\, 3/15, 2/15,\, -4/15] $
                         & $[0\degree,\, 78.46\degree,\, 82.34\degree,\, 105.47\degree]$ \\
        $-\pi\, [1,\, 3/15, -8/15,\, -14/15] $
                         & $[0\degree,\, 78.46\degree,\, 122.23\degree,\, 158.96\degree]$ \\
        \bottomrule
      \end{tabular*}
      }
  \end{table}
\end{example}

\begin{example}\label{ex:LargeArray}
  Consider the linear array $\br = (0,1,2,3,4,5,6,7,8,9,10,12)$, which has
  $N=12$ SSYTs. There are 14 partitions of $N=12$ not using the numbers 1
  and 4, namely
  \begin{align*}
    &\scriptstyle 2+2+2+2+2+2 = 2+2+2+3+3 = 3+3+3+3\\
    &\scriptstyle = 2+2+3+5 = 2+5+5 = 2+2+2+6 = 3+3+6\\
    &\scriptstyle = 6+6 = 2+3+7 = 5+7 = 2+2+8 = 3+9 = 2+10 = 12.
  \end{align*}
  The partitions $2+3+7$ and $12$ reached the timelimit of 450\,000 seconds,
  so that for these partitions we possibly have enumerated only a subset of
  all feasible solutions. For a subset of the remaining partitions
  all ambiguities found by enumerating feasible solutions of the
  corresponding smaller MIP are displayed in Table~\ref{tab:BigArrayAmbs},
  expressed as electrical angles. Here,~$\bof_j(v)$ and $\bof^{(k)}_6(v)$
  are vectors of electrical angles depending on a parameter, which are
  defined in Table~\ref{tab:fDef}. Thus, the (infinite) classes of
  ambiguities are given by combining the specified vectors of electrical
  angles into one large vector of~12 electrical angles depending on a
  number of parameters.

  \begin{table}[tb]
    \caption{Definition of $\bof_j(v)$ and $\bof^{(k)}_6(v)$ using electrical angles.}
    \label{tab:fDef}
    \centering
    {\tiny
      \begin{tabular*}{1.0\linewidth}{@{}l@{\extracolsep{\fill}}l@{}}
        \toprule 
        $\bof_2 = \pi\, [v - 1,\, v]$, & $v \in (0,1)$, \\
        $\bof_3 = \pi\, [v - 1,\, v - \tfrac{1}{3},\, v + \tfrac{1}{3}]$, & $v \in (0,\tfrac{2}{3})$, \\
        $\bof_5 = \pi\, [v - 1,\, v - \tfrac{3}{5},\, v - \tfrac{1}{5},\, v + \tfrac{1}{5},\, v + \tfrac{3}{5}]$, & $v \in (0,\tfrac{2}{5})$, \\
        $\bof_6^{(1)} = \pi\, [v - 1,\, v - \tfrac{14}{15},\, v - \tfrac{4}{15},\, v - \tfrac{1}{5},\,  v + \tfrac{1}{5},\,  v + \tfrac{3}{5}]  $, & $v \in (0,\tfrac{2}{5})$, \\
        $\bof_6^{(2)} = \pi\, [v - 1,\, v - \tfrac{3}{5},\,   v - \tfrac{8}{15},\, v + \tfrac{2}{15},\, v + \tfrac{1}{5},\,  v + \tfrac{3}{5}]  $, & $v \in (0,\tfrac{2}{5})$, \\
        $\bof_6^{(3)} = \pi\, [v - 1,\, v - \tfrac{3}{5},\,   v - \tfrac{1}{5} ,\, v - \tfrac{2}{15},\, v + \tfrac{8}{15},\, v + \tfrac{3}{5}]  $, & $v \in (0,\tfrac{2}{5})$, \\
        $\bof_6^{(4)} = \pi\, [v - 1,\, v - \tfrac{3}{5},\,   v - \tfrac{1}{5} ,\, v + \tfrac{1}{5},\,  v + \tfrac{4}{15},\, v + \tfrac{14}{15}]$, & $v \in (0,\tfrac{1}{15})$, \\
        $\bof_6^{(5)} = \pi\, [v - 1,\, v - \tfrac{3}{5},\,   v - \tfrac{4}{15},\, v + \tfrac{2}{15},\, v + \tfrac{8}{15},\, v + \tfrac{14}{15}]$, & $v \in (0,\tfrac{1}{15})$, \\
        $\bof_6^{(6)} = \pi\, [v - 1,\, v - \tfrac{14}{15},\, v - \tfrac{8}{15},\, v - \tfrac{2}{15},\, v + \tfrac{4}{15},\, v + \tfrac{1}{3}]  $, & $v \in (0,\tfrac{2}{3})$.\\
        \bottomrule
      \end{tabular*}
    }
  \end{table}

  In Table~\ref{tab:BigArrayStats} the number of feasible solutions, the
  time needed for enumerating the feasible solutions and the number of
  processed nodes in the enumeration process are displayed for each
  partition. Theorem~\ref{thm:uniformpartamb} yields one uniform ambiguity,
  namely
  \begin{align*}
    -\pi \cdot [\tfrac{5}{6},\, \tfrac{4}{6},\, \tfrac{3}{6},\,
    \tfrac{2}{6},\, \tfrac{1}{6},\, \tfrac{0}{6},\,
    -\tfrac{1}{6},
     -\tfrac{2}{6},\, -\tfrac{3}{6},\,
    -\tfrac{4}{6},\, -\tfrac{5}{6}]\T
  \end{align*}
  which is contained in the class
  \[
    [\bof_2(0),\, \bof_2(v_1),\, \bof_2(v_2),\, \bof_2(v_3),\,
    \bof_2(v_4),\, \bof_2(v_5)]
  \]
  of partition $2+2+2+2+2+2$.

  \begin{table}[tb]
    \caption{All found ambiguities in the linear array
      $\br = (0,1,2,3,4,5,6,7,8,9,10,12)$, for a subset of all possible
      partitions, expressed in electrical angles.}
    \label{tab:BigArrayAmbs}%
    {\tiny%
      \begin{tabularx}{\linewidth}{@{}X@{}}\toprule 
        \multicolumn{1}{c}{Partition $ 2+2+2+2+2+2$ --- one class of infinitely many ambiguities} \\
        \midrule
        \multicolumn{1}{@{}l}{$[\bof_2(0),\, \bof_2(v_1),\, \bof_2(v_2),\, \bof_2(v_3),\,
        \bof_2(v_4),\, \bof_2(v_5)] $,} \\
        \multicolumn{1}{c}{with $0 < v_1 < v_2 < v_3 < v_4 < v_5 < \tfrac{1}{2}$.} \\
        \midrule
        \multicolumn{1}{c}{Partition $2+2+2+3+3$ --- two classes of infinitely many ambiguities} \\
        \midrule
        \multicolumn{1}{@{}l}{$[\bof_2(0),\, \bof_2(v_1),\, \bof_2(v_2),\, \bof_3(w_1),\,
        \bof_3(w_2)] $,} \\
        \multicolumn{1}{c}{with $0 < v_1 < v_2 < \tfrac{1}{2}$, and $0 < w_1 < w_2 < \tfrac{1}{3}$.} \\
        \multicolumn{1}{@{}l}{$[\bof_3(0),\, \bof_2(v_1),\, \bof_2(v_2),\, \bof_2(v_3),\,
        \bof_3(w_1)] $,} \\
        \multicolumn{1}{c}{with $0 < v_1 < v_2 < v_3< \tfrac{1}{2}$, and $0 < w_1 < \tfrac{1}{3}$.} \\
        \midrule
        \multicolumn{1}{c}{Partition $3+3+3+3$ --- one class of infinitely many ambiguities} \\
        \midrule
        $[\bof_3(0),\, \bof_3(v_1),\, \bof_3(v_2),\, \bof_3(v_3)] $, with $0 < v_1 < v_2 < v_3 < \tfrac{1}{3}$. \\
        \midrule
        \multicolumn{1}{c}{Partition $2+2+3+5$ --- three classes of infinitely many ambiguities} \\
        \midrule
        $[\bof_3(0),\, \bof_2(v_1),\, \bof_2(v_2),\, \bof_5(w)] $, with $0 < v_1 < v_2 < \tfrac{1}{2}$ and $0 < w < \tfrac{1}{5}$. \\
        $[\bof_2(0),\, \bof_2(v),\, \bof_3(w),\, \bof_5(u)] $, with $0 < v< \tfrac{1}{2}$, $0 < w < \tfrac{1}{3}$ and $0 < u < \tfrac{1}{3}$. \\
        $[\bof_5(0),\, \bof_2(v_1),\, \bof_2(v_2),\, \bof_3(w)] $, with $0 < v_1 < v_2 < \tfrac{1}{2}$ and $0 < w < \tfrac{1}{3}$. \\
        \midrule
        \multicolumn{1}{c}{Partition $2+5+5$ --- two classes of infinitely many ambiguities} \\
        \midrule
        $[\bof_2(0),\, \bof_5(v_1),\, \bof_5(v_2)] $, with $0 < v_1 < v_2 < \tfrac{1}{5}$. \\
        $[\bof_5(0),\, \bof_2(v),\, \bof_5(w)] $, with $0 < v < \tfrac{1}{2}$ and $0 < w < \tfrac{1}{5}$. \\
        \midrule
        \multicolumn{1}{c}{Partition $2+2+2+6$ --- twelve classes of infinitely many ambiguities} \\
        \midrule
        $[\bof_2(0),\, \bof_2(v_1),\, \bof_2(v_2),\, \bof_6^{(k)}(w)] $, with $0 < v_1 < v_2 < \tfrac{1}{2}$, $k \in [6]$. \\
        $[\bof_6^{(k)}(0),\, \bof_2(v_1),\, \bof_2(v_2),\, \bof_2(v_3)] $, with $0 < v_1 < v_2 < v_3 < \tfrac{1}{2}$, $k \in [6]$. \\
        \midrule
        \multicolumn{1}{c}{Partition $3+3+6$ --- twelve classes of infinitely many ambiguities} \\
        \midrule
        $[\bof_3(0),\, \bof_3(v),\, \bof_6^{(k)}(w)] $, with $0 < v < \tfrac{1}{2}$, $k \in [6]$. \\
        $[\bof_6^{(k)}(0),\, \bof_3(v_1),\, \bof_3(v_2)] $, with $0 < v_1 < v_2 < \tfrac{1}{3}$, $k \in [6]$. \\
        \midrule
        \multicolumn{1}{c}{Partition $6+6$ --- thirty-six classes of infinitely many ambiguities} \\
        \midrule
        $[\bof_6^{k_1}(0),\, \bof_6^{(k_2)}(w)] $, with $k_1$, $k_2 \in [6]$. \\\bottomrule
      \end{tabularx}
    }
  \end{table}

  \begin{table}[tb]
    \caption{Number of solutions, solution times and number of processed
      nodes for the MIPs for each partition for the linear array
      $\br = (0,1,2,3,4,5,6,7,8,9,10,12)$.}
    \label{tab:BigArrayStats}
    \centering
    {\footnotesize
    \begin{tabular*}{1.0\linewidth}{@{}l@{\extracolsep{\fill}}rrr@{}}
      \toprule
      Partition   & \# Solutions & Time (s) & \# Nodes \\
      \midrule
      2+2+2+2+2+2 & \num{  160} &       \num{   719.43} &       \num{  2204798} \\
      2+2+2+3+3   & \num{26664} &       \num{189691.62} &       \num{ 74705045} \\
      3+3+3+3     & \num{  405} &       \num{  5790.55} &       \num{ 17067233} \\
      2+2+3+5     & \num{42336} &       \num{300709.50} &       \num{102050134} \\
      2+5+5       & \num{ 1680} &       \num{ 20150.43} &       \num{ 45847950} \\
      2+2+2+6     & \num{ 9968} &       \num{ 85418.02} &       \num{ 60992369} \\
      3+3+6       & \num{ 2736} &       \num{ 26180.16} &       \num{ 56723306} \\
      6+6         & \num{  130} &       \num{  2495.70} &       \num{  9615581} \\
      2+3+7       & \num{13318} & \g \! \num{450000.00} & \g \! \num{157647076} \\
      5+7         & \num{ 1289} &       \num{ 37766.13} &       \num{ 68727902} \\
      2+2+8       & \num{ 8046} &       \num{260789.23} &       \num{144652756} \\
      3+9         & \num{ 1188} &       \num{ 44698.63} &       \num{ 60887227} \\
      2+10        & \num{ 1830} &       \num{ 75348.62} &       \num{ 83449067} \\
      12          & \num{  315} & \g \! \num{450000.00} & \g \! \num{ 70499108} \\
      \bottomrule
    \end{tabular*}
    }
  \end{table}
\end{example}

\section{Conclusion} \label{sec:Conclusion}

We demonstrated that for several integer linear arrays our method is able to find more ambiguities than were
known previously by the methods from Manikas and
Proukakis~\cite{ManikasP98} (general case) and the methods from
Dowlut~\cite{Dowlut02} and Manikas~\cite{Manikas04} (symmetric
case). It turns out that arrays with a small number of SSYTs also have a
small number of (infinite) classes of ambiguities and the number of
SSYTs increases with the number of holes in the corresponding linear
array.

Example~\ref{ex:InfiniteSolutions} shows that at least for
symmetric linear arrays there exist non-uniform ambiguities that cannot be
found with our approach described in Section~\ref{sec:EnumMVS}. Therefore,
the interesting question arises, whether there can exist ambiguities in
non-symmetric linear arrays that cannot be represented using sums of
rotated minimal vanishing sums
(c.f. Example~\ref{ex:LimitationOfApproach} and the discussion
thereafter).

\bibliographystyle{abbrv}
\bibliography{uniqueLocalization}

\end{document}